\documentclass[12pt]{article}
\newcommand{\av}{\overline}

\makeatletter
\newcommand{\di}{\displaystyle}
\renewcommand{\dh}{\hspace{-2 mm}\di}

\begin{document}
\begin{center}
{\Large {\bf DYNAMICAL SCREENING OF GRAVITATIONAL INTERACTION AND
PLANETARY MOTIONS IN MODIFIED SOLAR POTENTIAL}}\\ \large {A.G.
Bashkirov\footnote{Corresponding author.
 {\it E-mail
address:} abas@idg.chph.ras.ru}, A.V. Vityazev and G.V.
Pechernikova}\\ Institute for Dynamics of Geospheres RAS, \\
Leninskiy prosp. 38 (bldg.6), 117334 Moscow, Russia\\
\end{center}
\begin{abstract}
A density disturbance in a system of gravitating mass, induced by
a moving selected body gives rise to a dynamical screening of
Newtonian potential of this body. When applied to the solar
planetary system it means that as a result of the motion of the
Sun in the Galaxy its effective force potential appears more weak
than the Newtonian potential. The relevant modifications of main
relations of the solar dynamics are considered here and it is
found in particular that the reestimated period of the Earth
revolution around the Sun rises in 1 second per year and semimajor
axis of the Earth orbit increases on 4 kilometers. Similar
relations are obtained for other planets too. It may be supposed
that the inclusion of these effects can help to explain the
observable anomalous acceleration of spacecrafts Pioneer 10 and
11.
\\
\end{abstract}

\section{Introduction}

It is a matter of general experience that a resting test charge in
a plasma produces the Debye screened potential. When  this charge
is moving the static screening decreases [1] and the potential
becomes anisotropic [2]. These properties are due to collective
effects in systems of charge particles. Here the similar effects
for systems of gravitating particles are discussed. For simplicity
we shall restrict our consideration to the model steady state
system of a homogeneous Maxwellian gas of gravitating bodies.

The problem of motion of the test star in a fluctuating force
field of another stars was first studied by Chandrasekhar [3] and
discussed later by Marochnik [4] using methods of kinetic gas
theory as an initial problem for a distribution function
describing the test star. In contrast to his approach, here we
consider a perturbation of the equilibrium state of a gas of
gravitating bodies as a response to a force field of the test
moving body following to our general approach developed in refs.
[5-7].

The main advantage of our approach is a derivation in Sec.2 of the
expression for a renormalized (effective) potential of the Sun and
calculation of its difference from the Newtonian solar potential
for planets of the solar system.

Next we calculate in Sec.3 corresponding corrections to the
periods and semimajor axis of revolutions of the Earth and other
planets.

\section{Gravitational screening}

In contrast to plasma or electrolyte where the screening of the Coulomb
potential of a charge particle is presumably due to the presence of
particles of an opposite charge and takes a place both for the moving
test particle (dynamical screening) and for the fixed one (static
Debye--H\"uckel screening) the screening of the Newtonian potential
of a selected test body in a system of gravitating bodies is possible
for the moving test body only.

In the model discussed below the Galaxy is considered (neglecting the dark
matter) as an equilibrium gravitating system of stars with a chaotic
distribution of velocities described by the Maxwellian
$f_0 ({\bf v}) = \rho_0(2\pi \tilde{v}^2)^{-3/2}\exp\{-v^2/2\tilde{v}^2\}$,
where $\rho_0$ is the density of stellar matter and $\tilde{v}$
velocities dispersion. The Sun is excluded from the whole system and considered
as the additional gravitating body of mass $M_{\odot }$ inserted at the
instant $t_0$ at the point ${\bf r_0=0}$ with the determined steady velocity
${\bf u}$. Introducing the test particle gives rise to a perturbation
$f_1({\bf r}, {\bf v},t) $ of the
system, so its distribution function takes the form
$f({\bf r}, {\bf v},t)=f_0({\bf v}) + f_1({\bf r}, {\bf v},t) $.

Then, in the linear approximation in the perturbation the system is described
by the Vlasov equation
\begin{equation}
\frac{\partial f_1({\bf r}, {\bf v},t)}{\partial t} + {\bf v}
\frac{\partial f_1({\bf r}, {\bf v},t)}{\partial {\bf r}}
 + \nabla\,\Phi\,\cdot\frac{\partial f_0({\bf r}, {\bf v},t)}{\partial {\bf v}}
 = 0.
\end{equation}
This is supplemented by the Poisson equation for the effective potential
\begin{equation}
\nabla^2\, \Phi = -4\pi Gm \int d^3 v f_1({\bf r}, {\bf v},t)-4\pi
G M_{\odot} \delta({\bf r} - {\bf u}\,(t-t_0)).
\end{equation}
We are interested in a steady--state behavior of the gravitating
system. It means physically that processes are considered at times
$t>\tau$, where $\tau=1/\sqrt {4\pi G \rho_0}$ is the
characteristic relaxation time. For our Galaxy (where $\rho_0=
1.9\cdot 10^{-23}\,{\rm g\,cm^{-3}} $) it is of order $10^7$ years
what falls far short of the time of existence of the solar system
(in the opposite case transition processes would be to take into
consideration). To exclude transition processes, we put $t_0 \to
-\infty$. Then the formal solution to equations (1) and (2) is
\begin{equation}
\nabla^2\, \Phi = 4\pi G m \int d^3\, v
\int^t_{-\infty}dt'\,\nabla \Phi ({\bf r}-{\bf v}(t-t'),t')\cdot
\frac{\partial f_0}{\partial {\bf v}} -4\pi G M_{\odot}
\delta({\bf r} - {\bf u}\,t)
\end{equation}
or in the Fourier transform:
\begin{equation}
[k^2 -4\pi G m \int d^3v\, \frac {{\bf
k}\cdot\partial{f_0}/\partial{\bf v}}{\omega -{\bf kv}+i\nu}] \Phi
({\bf k},\omega)=8\pi^2 GM_{\odot} \delta(\omega -{\bf ku}),
\end{equation}
whence
\begin{equation}
\Phi ({\bf k},\omega)=\frac {8\pi^2 GM_{\odot}}{k^2
\varepsilon_g({\bf k},\omega)} \delta(\omega -{\bf ku}).
\end{equation}
Here
\begin{equation}
\varepsilon_g({\bf k},\omega)=1-\frac{4\pi G}{k^2} \int d^3v\, \frac
{{\bf k}\cdot\partial
 f_0/\partial {\bf v}}{\omega -{\bf kv}+i\nu}
\end{equation}
is the gravitational permittivity of the system. This function
determines a reply of the gravitating system on a gravitational
disturbance. The concept of the gravitational permittivity is
well--known in the theory of gravitating system (see, e.g. [4, 8,
9]). Its zeroes define a dispersion equation of a gravitating
system of which complex roots determine a spectrum of elementary
excitations.

An infinitesimal imaginary value $i\nu $ is added to the frequency
$\omega$ in equations (4) and (6), which may be considered as a
result of introducing a "shadow" of the Boltzmann collision
integral $-\nu f_1$ into the r.h.s. of the Vlasov equation. Such
addition assures a selection of retarded solutions and accounts
for Landau damping in collisionless plasma. Then the permittivity
$\varepsilon_g({\bf k},\omega)$ becomes a complex function and can
be expressed as
\begin{equation}
\varepsilon_g({\bf k},\omega)= 1 -\frac{\kappa^2_J}{k^2}
W(z),\,\,\,\,\, z = \frac{\omega}{\sqrt{2}k\tilde{v}},
\end{equation}
where $\kappa_J=\left(4\pi G\rho_0/\tilde {v}^2\right)^{1/2}$  and
$W(z)=1+i(\pi)^{1/2} z\exp\{-z^2\}\,{\rm erfc}(-iz)$.

Calculating integrals of reverse Fourier transform
\begin{equation}
\Phi ({\bf r},t) =\frac {1}{(2\pi)^4}\int^\infty_{-\infty}d\omega\int d^3k\,
e^{\di -i(\omega t - {\bf k}{\bf r})}\, \Phi ({\bf k},\omega)
\end{equation}
we choose the axis $z$ along ${\bf u}$ and introduce the
dimentionless variables $Z=(z-ut)\kappa_J,\,\,X=x\kappa_J,\,\,
{\bf K}={\bf k}/\kappa_J,\,\,V=u/(\sqrt2\tilde{v})$. Then, taking
into account cylindrical symmetry we get
\begin{eqnarray}
\Phi ({\bf r},t)&\dh =&\dh\frac {GM_{\odot}\kappa_J}{2\pi^2
}\int^\pi_{-\pi}d\phi \int^\pi_0d\theta\int dK
K^2\sin\theta\frac{\exp\{iK(X\sin\theta\cos\phi+
Z\cos\theta)\}}{K^2-W(V\cos\theta)}=\nonumber\\ &\dh=&\dh
\frac{GM_{\odot}\kappa_J}{(X^2+Z^2)^{\frac 1{2}}}\left[1+
(X^2+Z^2)^{\frac 1{2}}I(X,Z,V) \right],
\end{eqnarray}
where
\begin{equation}
I(X,Z,V)=\frac {1}{2\pi^2 }\int^\pi_{-\pi}d\phi
\int^\pi_0d\theta\int dK\,W(V\cos\theta)\sin\theta\frac{\exp
\{iK(X\sin\theta\cos\phi+ Z\cos\theta)\}}{K^2-W(V\cos\theta)}.
\end{equation}
This integral determines a deviation of the effective potential
from the Newtonian one described by the first term in the square
bracket of the equation (9). After taking the Cauchy integral over
$K$ we find
\begin{eqnarray}
I(X,Z,V) &\dh=&\dh\frac {1}{4\pi^2}\int^\pi_{-\pi}d\phi
\int^\pi_0d\theta \sin\theta \sqrt{-W}\,
\Bigl[-2\,i\,\sinh ({\Delta}{\sqrt{- W}}){\rm Ci}
(-i\,{\Delta}{\sqrt{-W}})+ \Bigr.\nonumber\\
&\dh {}&\dh + \Bigl.\cosh ({\Delta}{\sqrt{- W}})
\,\left(\pi + 2\,i \,{\rm Shi}({\Delta}
{\sqrt{-W}})\right)\Bigr] ,
\end{eqnarray}
where $\Delta = X\sin\theta\cos\phi+ Z\cos\theta $.
To calculate numerically values of this integral over an arbitrary range
of variations of the parameters
$V$ and $Z$ should present no problems but for the discussed subject of
planetary motion in the solar system we can confine our consideration to
concrete values of these parameters.

For stellar (barionic) matter in the Galaxy we have
$\kappa_J\simeq 2.6\cdot10^{-21}\,\rm cm^{-1}$, thus the values
$Z$ and $X$ can be estimated as $10^{-9}<|Z|<10^{-6}$,
$10^{-9}<|X|<10^{-6}$ for planets of the solar system of which
orbits remote from the Sun at distances $10^{12}<r<10^{15}\,\rm
cm$. Since a peculiar velocity of the Sun in the Galaxy is
$1.95\cdot 10^6\,\rm cm/c$, and the velocity dispersion in the
Galaxy is $\tilde{v}\simeq 1.55\cdot 10^6\,\rm cm/c$, we put
$V=1$.

As a result of calculation of the integral (11) at $V=1$ we get
$I=-0.3677 $. When varying $V$ within the range from 0.3 to 3.0,
the numerical value of the integral $I $ varies from -0.3 to
-0.47.

\section{Correction to revolution periods}
According to the equation (9) and obtained numerical estimation of the
integral $I$ the energy of the renormalized gravitational interaction between
the Earth and the Sun can be represented as
\begin{equation}
U(R)=U_0+\delta U,\,\,\,U_0=-A/r,\,\,\,\delta U=A\gamma,\,\,\,
A=GM_{\odot }M_{\oplus },\,\,\,\gamma= - I \kappa_J.
\end{equation}
Thus, in such approximation the correction $\delta U $ to the Newtonian
energy of interaction is constant value independent on $r$.

Before proceeding to a calculation of its influence on the Kepler
period of the Earth revolution it is necessary to check whether it disturb
the closure of the finite trajectory of the Earth around the Sun.

In the classical two--body problem (after its reduction to the
problem of the one body motion in a central force field) an angle
of precession of the perihelion $\Delta \varphi$ (see, e.g. [10])
is defined as
\begin{equation}
\Delta \varphi =2\int^{r_{max}}_{r_{min}}\frac{\frac {M} {r^2} dr}
{\sqrt {2m(E-U_0-\delta U)-\frac{M^2}{r^2}}},
\end{equation}
where $r_{max}$ and $r_{min}$ are the maximum and minimum of the
radius--vector $r$, $m=\frac{M_{\odot}M_{\oplus }}{M_{\odot
}+M_{\oplus }}$ reduced mass, $$M=mr^2\dot\varphi,\,\,\,
E=\frac{m\dot r^2}{2}+\frac{M^2}{2mr^2} + U.$$ The last two
values, the moment $M$ and energy $E$, are the motion integrals.

The condition of closure of a trajectory is that the angle of
precision to be of the form $\Delta \varphi =2\pi k/n$ where $k,n$
are integers. For Newtonian potential this condition is fulfilled
($\Delta \varphi=2\pi $).

As a result of expanding of the integrand of the equation (13) in
$\delta U $ the zeroth member gives $2\pi $, and the first member
determines an additional precession of the perihelion as
\begin{equation}
\delta \varphi =\frac {\partial} {\partial M}
\left(\frac{2m}M\int_0^\pi r^2\delta U d\varphi \right)_0,
\end{equation}
where the expression within the brackets $(...)_0$ is estimated for the
unperturbed motion. With the use of the parameter $p$ and eccentricity $e$ of
an elliptic closed orbit we have
\begin{equation}
r=\frac{p}{(1+e\cos \varphi )},\,\,\,p=\frac{M^2}{mA},\,\,\,
e=\sqrt{1+\frac{2EM^2}{mA^2}}.
\end{equation}
Then
\begin{eqnarray*}
\delta \varphi  &=& \frac{2\gamma }{mA}\frac {\partial}{\partial
M} \left(M^3\int_0^\pi\frac{d\varphi }{(1+e\cos \varphi
)^2}\right) \\ &=&\frac{2\gamma }{mA}\frac {\partial}{\partial M}
(\frac{M^3\pi }{(1-e^2)^{3/2}}) =\frac{2\gamma \pi }{mA}\frac
{\partial}{\partial M}
\left(\frac{M^3}{(\frac{2|E|M^2}{mA^2})^{3/2}}\right)=0.
\end{eqnarray*}
Therefore, the perturbed trajectory remains closed.

We estimate now a change of the revolution period $T$ under
influence of the perturbation to the energy $\delta U$. It is not
difficult to check that all the mathematical treatment leading
[10] to the third Kepler law for undisturbed Newtonian potential
\begin{equation}
T_0=\pi A\sqrt{\frac{m}{2|E_0|^3}},\,\,\,
E_0=\frac{m\dot r^2}2+\frac{M^2}{2mr^2}+U_0(r)
\end{equation}
remains valid when we change $E_0$ for $E=E_0+\delta U$ and leads to the
modified form
\begin{equation}
T=\pi A\sqrt{\frac{m}{2|E_0+\delta U|^3}}=\pi A\sqrt{\frac
m{2|E_0|^3}}\left(1+\frac{\delta U}{E_0}\right)^{-3/2}\simeq
T_0\left(1+\frac{3A\gamma }{2|E_0|}\right),
\end{equation}
whence
\begin{equation}
\frac{T-T_0}{T_0}=\frac{3A\gamma }{2|E_0|}.
\end{equation}
It can be noted that $a_0=\frac A{2|E_0|}$ is the semimajor axis.
Substituting here the value of $a_0$ we get
\begin{eqnarray}
\frac{T-T_0}{T_0}= 3\gamma a_0  &=& 4.25\cdot
10^{-8}\left[\frac{I}{-0.3677}\right] \left[\frac{G}{6.67\cdot
10^{-8}}\right]^{1/2} \left[\frac{\rho_0}{1.9\cdot
10^{-23}}\right]^{1/2}\times \nonumber \\ &{}&\times
\left[\frac{\tilde v}{1.55\cdot 10^6}
\right]^{-1}\left[\frac{a_0}{1.496\cdot 10^{13} }\right]
\end{eqnarray}
which corresponds to $\Delta T_{\oplus } \simeq 1.3$ sec/year (1
year $\simeq 3.15\cdot 10^{7}$ sec)

Thus, accounts for screening of the Newtonian potential gives rise to
lengthen of the ET year on 1 second approximately what compensate
(in the order of the value) the systematic difference
$\av{\Delta T} =\av{UT-ET}=\av{UT} -{ TAI}\simeq 0.8$ sec.

Another consequence of the renormalization of the Newtonian
potential is modification of parameters of the Earth orbit. In
particular, the Earth semimajor axis $a_{\oplus }\propto |E|^{-1}$
increase on the value
\begin{equation}
\frac {\Delta a_{\oplus }}{a_{\oplus }}\simeq \frac {A\gamma
}{|E_0|}\simeq 2.83\cdot 10^{-8},
\end{equation}
that is, $\Delta a_{\oplus }\simeq 4.25\cdot 10^5\,$cm.

Similar calculations may be performed for other planets of the
solar system. Their results are represented in the Table.

\begin{center}
{\renewcommand{\arraystretch}{0}%
\begin{tabular}{|l||c|c|c|c|}
 \hline
 \strut Planet & $\Delta T/T$ &$\Delta T,\,\,sec$ & $\Delta a/a$ & $\Delta a, \,\,$ km\\
 \hline
 \rule{0pt}{2pt}&&&&\\
 \hline
 \strut {\it  Venus} &$ 3.1\cdot 10^{-8}$&$ 0.6 $&$ 2.1\cdot 10^{-8}$& 2.2 \\
 \hline
 \strut {\it  Earth} &$ 4.3\cdot 10^{-8}$&$ 1.3 $&$ 2.8\cdot 10^{-8}$& 4.2 \\
 \hline
 \strut {\it  Mars} &$ 6.5\cdot 10^{-8}$&$ 3.8 $&$ 4.3\cdot 10^{-8}$& 9.9 \\
 \hline
 \strut {\it  Jupiter} &$ 2.2\cdot 10^{-7}$&$ 0.83\cdot 10^{2} $&$ 1.5\cdot 10^{-7}$&$ 1.1\cdot 10^{2}$\\
 \hline
 \strut {\it  Saturn} &$ 4.1\cdot 10^{-7}$&$ 3.8\cdot 10^{2} $&$ 2.7\cdot 10^{-7}$&$ 3.9\cdot 10^{2}$\\
 \hline
 \strut {\it  Uranus} &$ 8.2\cdot 10^{-7}$&$ 2.2\cdot 10^{3} $&$ 5.4\cdot 10^{-7}$& $1.6\cdot 10^{3}$\\
 \hline
 \strut {\it  Neptune} &$ 1.3\cdot 10^{-6}$&$ 6.7\cdot 10^{3} $&$ 8.5\cdot 10^{-7}$&$ 3.8\cdot 10^{3}$\\
 \hline
\end{tabular}}
\end{center}

Thus, we see that relative time and space corrections to the far
giant planets motion resulted from the renormalization of the
solar potential are of order $10^{-6}$. It may be supposed that
they can be detected with the use of spacecrafts on far solar
orbits.

The above results are based on some approximations. The most
important of them are our choice of galactic disk volume density
$\rho_0 = 1.9\cdot 10^{-23}\,{\rm g\,cm^{-3}}$ having an influence
on the value of the Jeans wave number $\kappa_J$ and total
ignoring of a dark matter. As it was shown in [6], inclusion of
the dark matter gives rise to a sufficient decrease of an
effective Jeans wave number and enhancing of the dynamical
screening. Therefore, an experimental checking of the above
results could give an additional information related to the dark
matter in the Galaxy.

It would be interesting to apply our results to discussion of the
observed [11, 12] anomalous Pioneer 10 and 11 acceleration with a
magnitude $\sim 8\times 10^{-8}\,{\rm cm\,s^{-2}}$. Its relative
value is of order of $10^{-4}$, that is much greater our
corrections to orbits of the giant planets being at the same
distance ($20\div 30\,$AU) from the Sun. An expected value of the
spacecraft acceleration was calculated with the use of two
independent programs, that are JPL's Orbit Determination Program
(ODP) and Aerospace Corporation's Compact High Accuracy Satellite
Motion Program (CHASMP). Both programs use the planetary efemeris,
and timing inputs. Without knowledge of the programs we can only
suppose that taking into account of our corrections to these data
can produce a change of the expected acceleration and, probably,
provide a reasonable explanation for the observed anomalous
acceleration.

\subsection*{References}
1.{\it Cooper, G.} 1969, Phys. Fluids, {\bf 12}, 2707\\[0pt]2.{\it
Trophymovich, E. $\&$ Kraynov, V.} 1993, JETP {\bf 104},
3971\\[0pt]3.{\it Chandrasekhar, S.} 1942, Principles of Stellar
Dynamics (Chicago: Chicago Univ. Press)\\[0pt]4.{\it
 Marochnik, L.S.} 1968, Sov. AJ, {\bf 11}, 873\\[0pt]
5.{\it Vityazev A.V., Bashkirov A.G.} 1996, in Dynamics,
Ephemerids and Astrometry of the

Solar System, ed. S.Ferraz-Mello, B.Morando $\&$ J.--E.Arlot
(Amsterdam:

Kluwer Ac.Publ.), 327\\ 6.{\it Bashkirov A.G., Vityazev A.V.} 1998
Aph.J. {\bf 497}, No.1, Part 1, 10\\ 7.{\it Bashkirov A.G.,
Vityazev A.V., Pechernikova G.V.} 2001, Doklady Russian Academy

Sciences, {\bf 377}, No.3\\  8.{\it Binney J., Tremaine S.}
Galactic Dynamics. Princeton, 1987\\ 9.{\it Saslaw, W.C.}
Gravitational Physics of Stellar and Galactic Systems (Cambridge:

Cambridge Univ. Press)  1985.\\10.{\it Landau L.D., Lifshits E.M.}
Classical Mechanics. Pergamon Press, 1965.\\11.{\it Anderson J.D.,
Laing P.A., Lau E.L., Liu A.S., Nieto M.M., Turyshev S.G.} 1998,
Phys.

Rev. Lett., {\bf 81}, 2858.\\12.{\it Turyshev S.G., Anderson J.D.,
Laing P.A., Lau E.L., Liu A.S., Nieto M.M., } 1999,

gr-qc/9903024.

\end{document}